\documentclass[%
	prl,
	aps,
	floatfix,
	showpacs,
	twocolumn,
]{revtex4}

\usepackage{bm}
\usepackage{amsmath}
\usepackage{amssymb}
\usepackage{latexsym}
\usepackage{amsfonts}

\usepackage{graphics}

\newcommand{\f}[1]{\mbox{\boldmath$#1$}}

\newcommand{\ket}[1]{|#1\rangle} 
\newcommand{\bra}[1]{\langle#1|} 
\newcommand{\bracket}[2]
{\langle#1|#2\rangle} 
\newcommand{\proj}[1]{\ket{#1}\bra{#1}}
\newcommand{\ERW}[1]{\langle #1 \rangle}
\newcommand{\COM}[2]{\left[ #1, #2 \right]}

\newcommand{\nn}{\nonumber\\} 
\newcommand{\order}[1]{ \ensuremath{ {\cal O}(#1) } }
\newcommand{\ord}{{\cal O}}

\newcommand{\abs}[1]{\left|#1\right|}

\newcommand{\ie}{\textit{i.e.}}

\newcommand{\etal}{\textit{et al.}}

\begin{document}

\title{Non-Markovian decoherence in the adiabatic quantum search algorithm}

\author{Markus Tiersch and Ralf Sch\"utzhold$^*$}

\affiliation{Institut f\"ur Theoretische Physik,
Technische Universit\"at Dresden, 01062 Dresden, Germany}

$^*$ email: {\tt schuetz@theory.phy.tu-dresden.de}

\begin{abstract}
We consider an adiabatic quantum algorithm (Grover's search routine)
weakly coupled to a rather general environment,
i.e.,  without using the Markov approximation.
Markovian errors generally require high-energy excitations
(of the reservoir) and tend to
destroy the scalability of the adiabatic quantum algorithm.
We find that, under appropriate conditions (such as low temperatures),
the low-energy (i.e., non-Markovian) modes of the bath are most important.
Hence the scalability of the adiabatic quantum algorithm depends on the 
infra-red behavior of the environment: a reasonably small 
coupling to the three-dimensional electromagnetic field does not 
destroy the scaling behavior, whereas phonons or localized degrees of 
freedom can be problematic. 
\end{abstract}


\maketitle

%

In contrast to (conventional) sequential quantum algorithms in which the
solution to a given problem is obtained by measuring a final state after
a sequence of operations (gates) acting on an initial state,
the underlying idea of adiabatic quantum computation is to encode that
solution in the ground state of a suitably designed Hamiltonian instead
\cite{Farhi}.
Since the usual relaxation into that ground state typically takes an
exponentially long time for a complicated and strongly coupled Hamiltonian
(as the system is very likely to be trapped in a local minimum),
the desired ground state is reached by exploiting the adiabatic theorem:
Given a time-dependent Hamiltonian $H(t)$ with its instantaneous energy
eigenbasis $H(t)\ket{E_k(t)}=E_k(t)\ket{E_k(t)}$, the actual quantum state
$\ket{\psi(t)}$ of the system stays near the ground state $\ket{E_0(t)}$
\begin{equation}
\label{eq:adiab_expansion}
\ket{\psi(t)}
\approx
\ket{E_0(t)}
+
\sum_{k>0}
\frac{\bra{E_k(t)} \dot H(t) \ket{E_0(t)}}{[E_k(t)-E_0(t)]^2}
\ket{E_k(t)}
\,,
\end{equation}
provided that the evolution is slow enough, i.e., that the first-order
corrections of this adiabatic expansion (second term on the r.h.s.) are
small.
Note that the dynamical phase 
$\varphi_k(t)=-i\int_0^t d\tau E_k(\tau)$
and the (geometrical) Berry phase
$\gamma_k(t)=i\int_0^t d\tau \bracket{E_k(\tau)}{\dot E_k(\tau)}$
have been absorbed into $\ket{E_k(t)}$ for brevity.
Therefore, by slowly evolving the quantum system from an initial Hamiltonian
with an easy to prepare ground state (the initial state) into the final
problem Hamiltonian, the system ends up in (or close to) the desired ground
state (encoding the sought-after solution).
For infinite running times, the system would perfectly stay in its ground
state as the non-adiabatic corrections become arbitrarily small.
But finite running times still allow for a reasonably high fidelity of the
final ground state -- the minimum runtime $T$ is then a measure for the
computational complexity of the algorithm.

%

One of the major motivations for adiabatic quantum computation
(and the main advantage in comparison to sequential quantum algorithms)
is the relative robustness of the ground state against decoherence
caused by the inevitable coupling to the environment --
which makes it a very promising candidate for an experimental realization.
However, so far there are only a few quantitative investigations of the
impact of decoherence on adiabatic quantum computers.
In addition, these investigations predominantly rely on master equations
\cite{Childs,Aaberg} or superoperators \cite{Sarandy}
(which are both based on the Markov approximation)
or classical (Markovian) noise \cite{Roland,Shenvi}.
But such processes require short time-scales and, in turn, large energy
scales -- which is a cause for concern that the Markov approximation might
not be suitable for adiabatic quantum computation, in which the system and
the environment are supposed to have a rather low temperature.
In order to bridge this gap, we consider an adiabatic quantum algorithm
(Grover's search routine) weakly coupled to a rather general environment
and calculate the impact of decoherence directly, i.e., without resorting
to the Markov approximation \cite{Note}. 

%

The quantum algorithm we are going to study is an adiabatic version of
Grover's search routine \cite{Grover} defined by the Hamiltonian
(acting on an $n$-qubit system)
\begin{equation}
\label{eq:grover_hamiltonian}
H(t) = \big[ 1-s(t) \big] \big( \openone - \proj{\psi_0} \big)
+ s(t) \big( \openone - \proj{w} \big)
\,,
\end{equation}
with the initial ground state
$\ket{\psi_0}=\sum_{x=0}^{N-1} \ket{x}/\sqrt{N}$
being the coherent superposition of all numbers $x$
and the final ground state
$\ket{w} \in \{\ket{x}\}_{x=0}^{N-1}$
denoting the problem's solution.
It solves the problem of finding a marked item $w$ in an unstructured
list of length $N=2^n$ and was first introduced in the framework of
sequential quantum computation \cite{Grover}.
Note that not the marked state $\ket{w}$ itself but only the projector
$\proj{w}$ onto this state is used in the Hamiltonian
(similar to a quantum oracle: not knowing the solution,
but being able to recognize it).
We choose the adiabatic quantum search algorithm because its parameters
such as the instantaneous energy eigenbasis and the fundamental energy gap
can be calculated analytically -- but we expect our results to reflect
the typical behavior of general avoided level crossings.

The time-dependence is governed by a strictly monotonic function $s(t)$
that interpolates between the initial and the final Hamiltonian with
$s(0)=0$ and $s(T)=1$, so that $T$ is the algorithm's running time.
The essential dynamics take place in a two dimensional subspace
spanned by \mbox{$\{\ket{w},\ket{\psi_0}\}$}.
After a Gram--Schmidt orthogonalization, the basis we use for our
calculations reads
\mbox{$\{\ket{w}, \ket{w^\perp}\}$} with
$\ket{w^\perp} = (\sqrt{N}\ket{\psi_0}-\ket{w})/\sqrt{N-1}$.
The fundamental (non-vanishing) energy gap between the ground state
and the first excited state
\begin{equation}
\label{eq:gap}
\Delta E
=
\sqrt{ 1 + 4 s(t) \big[ 1 - s(t) \big]
\left( \frac{1}{N} - 1 \right) }
\end{equation}
exhibits a minimum at $s=1/2$ of $\Delta E_\text{min}=1/\sqrt{N}$.
The total running time necessary for keeping the non-adiabatic corrections
small depends on the interpolation dynamics $s(t)$.
A constant velocity $\dot s=\rm const$ yields the same scaling
$T=\order{N}$ as the classical (brute-force) search,
cf.~Eq.~\eqref{eq:adiab_expansion}.
Adapted interpolations such as
\mbox{$\dot s \propto (\Delta E)^2$}
or
\mbox{$\dot s \propto \Delta E$},
on the other hand, recover the quadratic speed-up
$T=\order{\sqrt{N}}$ or $T=\order{\sqrt{N}\,\ln N}$,
respectively \cite{Schaller}.
We shall use these three specific cases of $s(t)$ for investigating
the impact of a coupling to an environment.

%

In order to describe the adiabatic quantum computer embedded in an
environment (open system), we use the following rather general
\emph{ansatz} for the Hamiltonian
\begin{equation}
H(t) = H_\text{sys}(t) + H_\text{env} + H_\text{int} 
\,,
\end{equation}
with $H_\text{sys}(t)$ denoting the explicitely time-dependent
Hamiltonian of the quantum computer (system) in
Eq.~(\ref{eq:grover_hamiltonian}),  $H_\text{env}$ describing the
(assumed to be stationary) dynamics of the environment
with $[H_\text{sys}(t),H_\text{env}]=0$, and $H_\text{int}$ the
weak interaction between them.
Switching to the interaction picture, the total density
matrix~$\rho$ (incorporating system plus environment)
evolves according to
\mbox{$\partial_t \rho(t) = -i \COM{H_\text{int}(t)}{\rho(t)}$}
(using units with $\hbar=1$).
For weak interactions between the system and the environment governed by
a small coupling constant $\lambda\ll1$,
it is quite natural to assume the following expansion of $H_\text{int}$
\begin{equation}
\label{H-int}
H_\text{int}
=
\lambda \sum_{a=1}^{n} \f{\sigma}_a \cdot \f{A}_a
+
\lambda^2 \sum_{a,b=1}^{n} \f{\sigma}_a \cdot \f{B}_{ab} \cdot \f{\sigma}_b
+
\order{\lambda^3}
\end{equation}
into one-qubit and two-qubits errors etc.
The $n$ qubits are labeled by $a$ and
$\f{\sigma}_a(t)=[\sigma_a^x(t),\sigma_a^y(t),\sigma_a^z(t)]$
is the vector of their Pauli matrices in interaction picture,
with the corresponding bath operators~$\f{A}_a(t)$ and $\f{B}_{ab}(t)$ etc.
In the following, we shall use perturbation theory in $\lambda$ and keep
only the lowest-order terms.
As we shall see below in Eq.~(\ref{perturbation}), this amounts to 
omitting the $\f{B}_{ab}$-term.
(Note, however, that this approximation does not imply the neglect of all
correlated errors.)

%

Initially the system of qubits has to be realized in (or close to)
the ground state~$\ket{\psi_0}$.
Thus, we assume that the initial full density operator is a direct product
\begin{equation}
\rho(0)
=
\rho_\text{sys}(0) \otimes \rho_\text{env}(0)
=
\proj{\psi_0} \otimes \rho_\text{env}(0)
\,,
\end{equation}
\ie, system and environment are not entangled at the beginning.
This should be a good approximation provided that the temperature
is much smaller than the initial energy gap and that the coupling
$H_\text{int}$ to the reservoir is weak.
When coupling an adiabatic quantum computer to an environment,
an essential quantity is the probability of measuring the
ground state at the end of the computation, at time~$T$, \ie,
whether or not the algorithm succeeded.
In the following, we shall assume perfect adiabatic evolution
(\ie, for $\lambda=0$, the success probability would be one)
and hence only consider perturbations due to the interaction
$H_\text{int}$, but not due to finite running time
(nor combinations of both).

It turns out that the final probability of measuring the first
excited state $\ERW{P_1(T)}$ is a necessary and sufficient
indicator for the success probability:
The spectrum of the Hamiltonian~(\ref{eq:grover_hamiltonian})
consists of the ground state $\ket{E_0(t)}$
and the first excited state $\ket{E_1(t)}$,
which come very close ($\Delta E_\text{min}=1/\sqrt{N}$)
at $s=1/2$, whereas all $N-2$ other states $\ket{E_{k>1}(t)}$
have the same constant energy and are therefore well separated
from the ground state by an energy gap of order one.
Since the temperature and hence the energies available in the
environment are supposed to be much smaller than that gap
of order one, transitions from the ground state to these states
$\ket{E_{k>1}(t)}$ are strongly (exponentially) suppressed
$\ERW{P_{k>1}(T)}\lll1$, where the operators
$P_k(T) = \proj{E_k(T)}$ are  projectors onto the
$k^\text{th}$ instantaneous energy eigenstate at the final
running time $T$.
The above observation that merely the ground state and the
first excited state in the vicinity of the avoided level crossing
(see below) are relevant supports our expectation that the results
derived in this special example~(\ref{eq:grover_hamiltonian})
represent characteristic features of general avoided level crossings.

%

The success of the algorithm then corresponds to $\ERW{P_0(T)} \approx 1$
or alternatively $\ERW{P_1(T)}\ll1$.
For a perfectly adiabatic evolution, the contributions to $\ERW{P_1(T)}$ 
of zeroth and first order in $\lambda$ vanish because of
$P_1(T) \approx \proj{\psi_0^\perp}$
and 
$\rho_\text{sys}(0) = \proj{\psi_0}$.
For the same reason, the $\f{B}_{ab}$-term in Eq.~(\ref{H-int}) do not 
contribute to lowest order $\order{\lambda^2}$. 
The second-order term in $\lambda$ gives
\begin{multline}
\label{perturbation}
\ERW{P_1(T)}
\approx
\lambda^2 \sum_{\substack{a,b=1\\ \mu,\nu=x,y,z}}^{n}
\int\limits_0^T dt_1 \int\limits_0^T dt_2\;
\ERW{A_a^\mu(t_1)A_b^\nu(t_2)}
\times
\\
\times
\bra{w^\perp}\sigma_a^\mu(t_1)\ket{w} \bra{w}\sigma_b^\nu(t_2)\ket{w^\perp}
\end{multline}
after symmetrizing the integrals and neglecting terms of order
$1/\sqrt{N}\lll1$.
This expression combines the system dynamics with the
environment's properties:
The environment correlation function is calculated with
$\rho_\text{env}(0)$ and the bath operators $A_a^\mu(t)$ and
the one-qubit operators have to be transferred to the interaction
picture
\begin{equation}
\sigma_a^\mu(t) = U_\text{sys}^\dag (t) \sigma_a^\mu U_\text{sys} (t)
\,,
\end{equation}
where $U_\text{sys} (t)$ is the unitary time evolution operator that is
implied by the system's Hamiltonian $H_\text{sys}(t)$ and can be
calculated with the known adiabatic expansion.
For this calculation, we used the leading term only, \ie,
perfect adiabatic evolution and in the large-$N$ limit.

Of the system matrix elements $\bra{w^\perp}\sigma_a^\mu(t)\ket{w}$
only those with $\mu=x,z$ contribute with order one,
the $\mu=y$ term is suppressed by a factor of $1/\sqrt{N}$.
So, for large $N$ the relevant matrix elements are
\begin{equation}
\bra{w^\perp}\sigma_a^x(t)\ket{w}
=
-e^{-i\int_0^td\tau \Delta E(\tau)}
\frac{1-s(t)}{\sqrt{N}\Delta E(t)}
\,,
\end{equation}
and the same for $\bra{w^\perp}\sigma_a^z(t)\ket{w}$ apart from an
additional sign $(-1)^{w_a+1}$, where $w_a$ is the $a^\text{th}$ bit of $w$,
\ie, the marked state $\ket{w}$ is an eigenstate of the operators
$\sigma_a^z$ with eigenvalues $(-1)^{w_a}$.
As one might expect, the matrix elements are strongly peaked at $s=1/2$,
\ie, at the point of the avoided level crossing.


For the estimation of the time integrals in Eq.~(\ref{perturbation}),
we need some information about the environment specifying the correlation
function $\ERW{A_a^\mu(t_1)A_b^\nu(t_2)}$.
In the Markov approximation, most of this information about the temporal
behavior of the reservoir is lost.
In order to compare our results with those obtained in the Markov
approximation, let us suppose that successive interactions are not
correlated (Markovian environment) and assume a correlation
function which is local in time
\begin{equation}
\ERW{A_a^\mu(t_1)A_b^\nu(t_2)}
=
A_{ab}^{\mu\nu} \delta(t_1-t_2)
\,.
\end{equation}
Insertion of this \emph{ansatz} into  Eq.~(\ref{perturbation}) yields
a failure probability of
\begin{equation}
\ERW{P_1(T)}
\propto
\lambda^2 \int_0^T dt
\left( \frac{1-s(t)}{\sqrt{N} \Delta E(t)} \right)^2
=
\order{\lambda^2 \sqrt{N}}
\end{equation}
for any of the three cases for $s(t)$ discussed after Eq.~(\ref{eq:gap}).
In this case, keeping the transition errors under control
$\ERW{P_1(T)}\ll1$ requires the coupling $\lambda$ to be exponentially
small with the number of qubits $\lambda=\order{1/\sqrt{N}}$, \ie,
the system of qubits has to be isolated exponentially good.
Experimentally, this is not feasible and would therefore render the
adiabatic quantum computer not scalable within the presence of such
a Markovian environment.
This result can be understood in the following way:
The Markov approximation is based on the assumption that the correlations
between system and reservoir due to their interaction at a certain point
of time spread very fast over the entire bath and hence can be neglected
for all interactions occurring later on.
However, these short time-scales require excitations of comparably high
energies.
Since the adiabatic quantum computer is in its ground state,
these high-energy excitations must have their origin in the bath.
On the other hand, a system coupled to such a reservoir containing
high-energy excitations for a relatively long time (the run-time of
adiabatic quantum computation strongly increases with the number of qubits)
is very liked to get excited -- \ie, the computation fails.


Instead of a Markovian bath we only assume a stationary
($[H_\text{env},\rho_\text{env}]=0$) reservoir
$\rho_\text{env}$ which allows for a Fourier decomposition of the
environment correlation function
\begin{equation}
\ERW{A_a^\mu(t_1)A_b^\nu(t_2)}
=
\int d\omega\, e^{-i\omega(t_1-t_2)} f_{ab}^{\mu\nu}(\omega)
\,,
\end{equation}
where $f_{ab}^{\mu\nu} (\omega)$ depends on the spectral distribution
(e.g., effective dimensionality) of the bath modes and the temperature etc.
This gives
\begin{multline}
\label{full}
\ERW{P_1(T)}
=
\lambda^2 \int d\omega\,
\sum_{a,b=1}^{n}
f_{ab}^{xx} (\omega)
\times
\\
\times
\abs{ \int_0^T dt \,
e^{ i[ \omega t + \int_0^td\tau \Delta E(\tau) ] }
\frac{1-s(t)}{\sqrt{N} \Delta E(t)} }^2
\end{multline}
plus similar terms including $f_{ab}^{xz}$, $f_{ab}^{zx}$,
and $f_{ab}^{zz}$ with the associated signs $(-1)$ and $(-1)^{w_a}$
for $x$ and $z$, respectively.
Note that these terms are suppressed (in comparison to $f_{ab}^{xx}$)
if all qubits are coupled to the same bath, for example, and if we assume
that the solution $w$ has (nearly) the same number of zeros and ones.

In order to evaluate the time integrations, it is useful to distinguish
three different domains of $\omega$: large positive frequencies
$\omega\gg\Delta E_\text{min}$, small frequencies
$\omega=\ord(\Delta E_\text{min})$, and large negative frequencies
$\omega\ll-\Delta E_\text{min}$.
For large frequencies $|\omega|\gg\Delta E_\text{min}$, the time integral
can calculated via the stationary-phase (or saddle-point) approximation.
The saddle points $t^*_\omega$ defined by a vanishing derivative of the 
exponent are at $\omega+\Delta E(t^*_\omega)=0$, which just corresponds 
to energy conservation.
Hence large positive frequencies $\omega\gg\Delta E_\text{min}$ do not
contribute at all -- which is quite natural since this corresponds to
a transfer of a large energy $\omega\gg\Delta E_\text{min}$ from the
system (which is in its ground state) to the reservoir.
However, since the temperature of the bath 
(and hence the typical energy of its excitations)
is typically much larger than $\Delta E_\text{min}$,
the opposite process is possible in general.
For small frequencies $\omega=\ord(\Delta E_\text{min})$, the
stationary-phase (or saddle-point) approximation cannot be applied
and energy conservation is also not a well-defined concept anymore
since the rate of the external time-dependence is just given by
$\ord(\Delta E_\text{min})$ and hence of the same order as the
energies under consideration.
In this case, we estimate the time-integral in (\ref{full}) by
omitting all phases (upper bound).
Altogether, we get
\begin{eqnarray}
\label{Altogether}
\ERW{P_1(T)}
&\sim&
\lambda^2 N
\int_{-\Delta E_\text{min}}^{+\Delta E_\text{min}}
d\omega\, f(\omega)
+
\nn
&&+
\frac{\pi\lambda^2}{2N}
\int_{\Delta E_\text{min}}^{1}
d\omega\,
\frac{f(-\omega)}{\omega^2 \dot s(t^*_\omega)}
\,,
\end{eqnarray}
where $f(\omega)$ is understood as the appropriate sum of the
$f_{ab}^{xx}$, $f_{ab}^{xz}$, $f_{ab}^{zx}$, and $f_{ab}^{zz}$
contributions.

In contrast to the first contribution, the second term depends on the
dynamics $s(t)$.
For the three scenarios ($\dot s=1/T$, $\dot s\propto\Delta E^2$,
or $\dot s\propto\Delta E/\sqrt{N}$) discussed after Eq.~(\ref{eq:gap}),
the second integrand scales as
$Nf(-\omega)/\omega^2$, $f(-\omega)/\omega^4$, or
$\sqrt{N}\,f(-\omega)/\omega^3$, respectively.
In all three cases, we see that the bath modes with large frequencies 
$|\omega|\gg\Delta E_\text{min}$ do not cause problems in
the large-$N$ limit:
The spectral function $f(-\omega)$ is supposed to decrease for large
$|\omega|$ as the bath does not contain excitations with large energies
by assumption (low temperatures).
Therefore, the potentially dangerous contributions stem from the
low-energy modes $\omega=\ord(\Delta E_\text{min})$ of the reservoir,
for which the Markov approximation is not applicable in general.
Independent of the dynamics $s(t)$, both, the first integral and
the lower limit of the second integral in Eq.~(\ref{Altogether})
yield the same order of magnitude
\begin{eqnarray}
\ERW{P_1(T)}
&\sim&
\lambda^2\,
\frac{f\left[\ord(\Delta E_\text{min})\right]}{\Delta E_\text{min}}
\,.
\end{eqnarray}
Since $\Delta E_\text{min}$ decreases as $1/\sqrt{N}$ in the
large-$N$ limit, the spectral function $f(\omega)$ must vanish in the
infra-red limit as $\omega$ or even faster in order to keep the error
$\ERW{P_1(T)}$ under control.
Exactly at the threshold $f(\omega)\sim\omega$, sub-leading contributions 
which scale polynomially (instead of exponentially) with the number of 
qubits $n\propto\ln\Delta E_\text{min}$ may become important:
Depending on the reservoir (e.g., all $\f{A}_a$ are the same versus all 
$\f{A}_a$ are independent), we get factors of $n^2$ or $n$ from the sum 
over all qubits and for $\dot s=1/T$ and $f(\omega)\sim\omega$, the 
lower limit of the integration yields $\ln\Delta E_\text{min}$. 


Let us exemplify tha above results by means of a simple model for the bath 
and consider the $n$ qubits to be encoded in a chain of $1/2$-spins fixed 
at the postions $\f{r}_a$, which are coupled to the magnetic field 
fluctuations $\f{B}(\f{r}_a)$ via their magnetic moment $\mu$
\begin{equation}
H_\text{int} = \mu \sum_{a=1}^n \f{\sigma}_a \cdot \f{B}(\f{r}_a)
\,.
\end{equation}
As the reservoir, we choose a thermal photon bath (whose temperature is
much smaller than the initial gap, but much bigger that the minimum gap)
in $D=1,2,3~$ spatial dimensions.
In this case, the spectral function can easily be evaluated to be
$f(\omega)\sim\omega^{D-1}$ by the usual normal-mode expansion of the 
electromagnetic field into creation and annihilation operators. 
Note that, at zero temperature, one would gain a factor of $\omega$
since the thermal particle content (Bose-Einstein distribution)
has a $1/\omega$ singularity in the infra-red (for a bosonic bath).
Consequently, a three-dimensional thermal photon bath does not
destroy the scalability of the adiabatic quantum algorithm under
consideration -- whereas such a reservoir restricted to one spatial
dimension (of infinite length) causes problems.
For $D=2$, the sub-leading terms scaling with poly$(n)$ become important. 

An entirely different situation arises when the qubits are coupled to
the amplitude of the field (representing the reservoir modes)  
instead of the field momentum density.
This might be the case for an effectively $D$-dimensional phonon bath 
and would result in $f(\omega)\sim\omega^{D-3}$. 
In such a situation, it is probably necessary to restrict the reservoir 
to a sufficiently finite volume (which generates an infra-red cut-off)
in order to keep errors small.
Note, however, that a finite volume would not improve the infra-red 
behavior of a spin bath consisting of localized degrees of freedom 
(e.g., nuclear spins). 

In summary, we examined the effect of decoherence on an adiabatic 
quantum computer by means of Grover's search algorithm weakly coupled 
to a rather general bath and found that (under reasonable conditions) 
the infra-red behavior of the reservoir limits the scalability of the 
algorithm.
Consequently, the Markov approximation does not capture the relevant 
features in this situation -- instead, the spectral function  
$f(\omega)$ of the bath provides a criterion to favor/disfavor 
certain physical implementations. 
Of course, a full analysis would have to include higher orders in 
$\lambda$ (see, e.g., \cite{Grifoni}). 
If the reservoir distributes the occupations equally between the two 
lowest levels (and there is only one level crossing), the algorithm 
still succeeds in half the cases.
On the other hand, if the bath modes get so strongly entangled with 
the state $\ket{\psi_0}$ (\ie, ``stick'' to it) that they prevent it from 
flipping to $\ket{w}$, the algorithm fails. 


This work was supported by the Emmy Noether Programme of the
German Research Foundation (DFG) under grant No.~SCHU~1557/1-1/2.
R.~S.~acknowledges fruitful discussions at the Banff workshop
``Spin, Charge and Topology in low D'' 2006 as well as support by the
Pacific Institute of Theoretical Physics.
M.~T.~thanks Gernot Schaller for helpful discussions.



\begin{thebibliography}{499}
\bibitem{Note}
After completing our derivation, we learned that M.~Amin \etal\,
have been pursuing a similar calculation, but with a different scope.
\end{thebibliography}
\end{document}